\documentclass[twocolumn,secnumarabic,amssymb, nobibnotes,aps,prl,superscriptaddress]{revtex4-2}
\usepackage{graphicx}
\usepackage{textcomp}
\usepackage{amsmath}
\usepackage{commath}
\usepackage{color}
\usepackage{appendix}
\usepackage{hyperref}

\usepackage{soul}

\usepackage{natbib}

\begin{document}

\title{Edge-selective reconfiguration in polarized lattices with magnet-enabled bistability}



\author{Luca Iorio}
\affiliation{Department of Civil, Environmental, and Geo- Engineering, University of Minnesota, Minneapolis, MN 55455, USA}
\affiliation{Department of Civil and Environmental Engineering, Politecnico di Milano, Milano, 20133, Italy}
\author{Raffaele Ardito}
\affiliation{Department of Civil and Environmental Engineering, Politecnico di Milano, Milano, 20133, Italy}
\author{Stefano Gonella}
\email{sgonella@umn.edu}
\affiliation{Department of Civil, Environmental, and Geo- Engineering, University of Minnesota, Minneapolis, MN 55455, USA}

\begin{abstract}

The signature topological feature of Maxwell lattices is their polarization, which manifests as an unbalance in stiffness between opposite edges of a finite domain. The manifestation of this asymmetry is especially dramatic in the case of soft lattices undergoing large nonlinear deformation under concentrated loads, where the excess of softness at the soft edge can result in the activation of sharp indentations. This study explores how this mechanical dichotomy between edges can be tuned and possibly extremized by working with soft magneto-mechanical metamaterials. The magneto-mechanical coupling is obtained by endowing the lattice sites with permanent magnets, which activate a network of magnetic forces that can interact with – either augmenting or competing with – the elasticity of the material. Specifically, under sufficiently large deformation that macroscopically alters the equilibrium positions of the sites, the attractive forces between the magnets can trigger bistable reconfiguration mechanisms. The strength of such mechanisms depends on the landscapes of elastic reaction forces exhibited by the edges, which are different due to the polarization, and is therefore inherently edge-selective. We show that, on the soft edge, the addition of magnets simply enhances the softness of the edge. In contrast, on the stiff edge, the magnets activate snapping mechanisms that locally reconfigure the cells and produce a lattice response reminiscent of plasticity, characterized by residual deformation that persists upon unloading.

\end{abstract}

\maketitle

\section{Introduction} \label{sec:Introduction}
Maxwell lattices are  critically coordinated mechanical lattices, i.e., lattices having an average coordination number (under periodic boundary conditions) equal to 2d, where d is the dimensionality of the space in which the lattices are embedded.
Topologically polarized Maxwell lattices, such as certain  kagome configurations, are known to display polarization, i.e., the ability to focus floppy modes (also known as mechanisms, i.e., deformation modes that do not involve the generation of stress in the bond)
on a given edge (i.e., boundary of a finite lattice domain), leaving the opposite edge rigid \cite{kane2014topological,rocklin2017directional,mao2018maxwell,baardink2018localizing,zhou2018topological}. In essence, polarization entails an excess of softness on one edge accompanied by a surplus of rigidity on the opposite edge. This property is 
protected by the k-space topology of the lattice band structure. Depending solely on the unit cell geometry and kinematics, it is an intrinsic property of the bulk which, however, manifests at the edges of a finite domain, according to the \textit{bulk-edge correspondence}. The topological protection implies that the polarization is preserved as long as the topology of the bulk is intact, and hence robust against anomalies (e.g., defects and disorder) that may exist on the edges \cite{zhang2018fracturing, paulose2015selective,chapuis2022mechanical,sun2012surface}.
When we load the soft edge 
with a point force, we tap into the local 
softness and induce deformation that localizes on the edge. 
In contrast, loading the rigid edge activates a rigid body motion \cite{charara2023cell}. These conditions are strictly predicted for ideal lattices, in which the bonds are harmonic springs and the sites act as perfect hinges that allow free rotations. This said, the behavior is observed, albeit diluted in strength, in topological kagome metamaterials, i.e., elastic continua shaped according to topological kagome configurations, in which the perfect hinges are replaced by elastic connections \cite{pishvar2020soft,zunker2021soft,paulose2015topological,zunker2021soft}. 

While polarization is a property of the linear elastic regime of deformation, its signature is observable deep into the nonlinear regime in the form of an asymmetry of the finite deformation patterns observed on opposite edges \cite{widstrand2024robustness,jolly2023soft}. Under a compressive load, the localized deformation of the floppy edge takes the form of a sharp indentation, 
promoted by the activation of large rotations 
of the cell triangles 
about their hinges, while the stiffness of the rigid edge prevents the onset of 
such mechanisms. 
Activating a geometric nonlinear response involving finite rotations in a \textit{structural metamaterial} requires operating with a soft material (e.g., an elastomer)~\cite{microtwist,zunker2021soft,pishvar2020soft,wu2024zero}. 
Therefore, a preliminary step of our investigation is to characterize 
experimentally the signature of polarization in a soft polarized kagome specimen 
through a qualitative and quantitative 
characterization of the differences in the edge deformation pattern observed between the edges. 
With the results of this assessment in hand, our key objective is to endow the lattice \textit{inter-cell interactions} with a new layer of \textit{functional complexity}, and study how this operation interplays with, and modifies, the asymmetry due to polarization, either \textit{quantitatively}, by accentuating the stiffness gap between the edges, or \textit{qualitatively}, by endowing the lattice with a new type of \textit{response asymmetry}. 

Specifically, we consider a soft (silicone rubber) kagome lattice whose cells are endowed with bistable mechanisms. The bistability is promoted by the magneto-mechanical coupling between the elasticity of the silicone and a superimposed network of discrete magnetic forces, obtained by placing permanent magnets at the centers of mass of the cell triangles, according to an appropriate sequence of attractive and repulsive pairs. 
Accordingly, a kagome cell has two equilibrium 
configurations, corresponding to two local minima of its strain energy landscape: the second stable configuration is achieved when, as a result of the relative rotation of the triangles, two adjacent magnets become sufficiently close that their attractive forces overcome the elastic reaction forces resulting from the bending of the ligaments. Under sufficiently large loads, selected cells can snap into the stable configuration, locally lifting the lattice periodicity. 
The utilization of magnets to harness previously unattainable dynamical ranges, transitions and general properties has been firmly established ~\cite{zhang2021tunable,nadkarni2016unidirectional,montgomery2021magneto}, but here a new class of topological structures is analyzed.
The practical way in which the bistability-promoting magnets are incorporated in the lattice is inspired by a prototype first put forth by Marshall and Ruzzene in~\cite{Ruzzene_mag}. In that work, whose main focus was a numerical assessment of the effects of cell reconfiguration on the phonon bands, this strategy was briefly demonstrated on a regular kagome prototype to illustrate how bistable mechanisms could potentially serve as configuration-switching tuning tools. Here, we embrace a similar technological solution and adapt it for polarized lattices, where the bistable mechanisms interplay non-trivially with the asymmetry provided by the polarization, with our focus directed explicitly towards the \textit{nonlinear deformation} regime involved in the actual reconfiguration process.

In classical problems involving bistable lattices, the bistable mechanisms are distributed uniformly across the domain~\cite{Damiano_1,Damiano_2} As a result, any induced reconfigurations typically affects the entire lattice, albeit often occurring in cascading sequences in the form of reconfigurational waves ~\cite{jiao2024phase,deng2017elastic} (with the exception of lattices in which a spatial modulation of the reconfiguration is introduced by design through defects~\cite{raney2016stable,Bertoldi_bistable,zhang2019programmable}).
In some cases bistable effects have been engineered to tap into new configurations, achieving reprogrammable metamaterials~\cite{pal2023programmable,wu2024zero,bilal2017bistable,xiu2022topological}. Here we hypothesize that the polarization introduces an asymmetry in the very way in which the bistable mechanisms are engaged at the edges, both in terms of size and shape of the regions where these mechanisms are established and in terms of the level of force under which they are triggered. Specifically, 
at the soft edge, where the deformation is large and localized in the neighborhood of the loading point, any local switching of the cells to their stable configuration will also remain largely confined to the loaded region.  
Moreover, since large levels of strain energy are stored in the hinges undergoing bending, the switching is heavily impeded, as the magnets struggle to overcome the restoring elastic forces resulting from the hinges, and is therefore set to revert upon unloading. In contrast, at the stiff edge, while it takes larger loads to trigger macroscopic deformation, 
when nonlinear deformation eventually occurs and bistable mechanisms are triggered, the switching effects are expected to leak deeper into the bulk, following the long-wavelength range of the very deformation patterns that trigger them. More importantly, the reconfiguration will be more stable upon unloading, since the magnetic forces can more easily overcome the restoring elastic forces from the hinges, leading to the onset of residual deformation. 
The remainder of this paper presents a series of experiments aimed at verifying these hypotheses and documenting the proliferating landscape of edge-selective scenarios that are enabled by the cooperative interplay between polarization and magnetic-enabled instabilities.  

\section{A simplified mechanistic perspective on polarization} \label{sec:top_teory_new}
\begin{figure*}
\centering
\includegraphics[width = 1\textwidth]{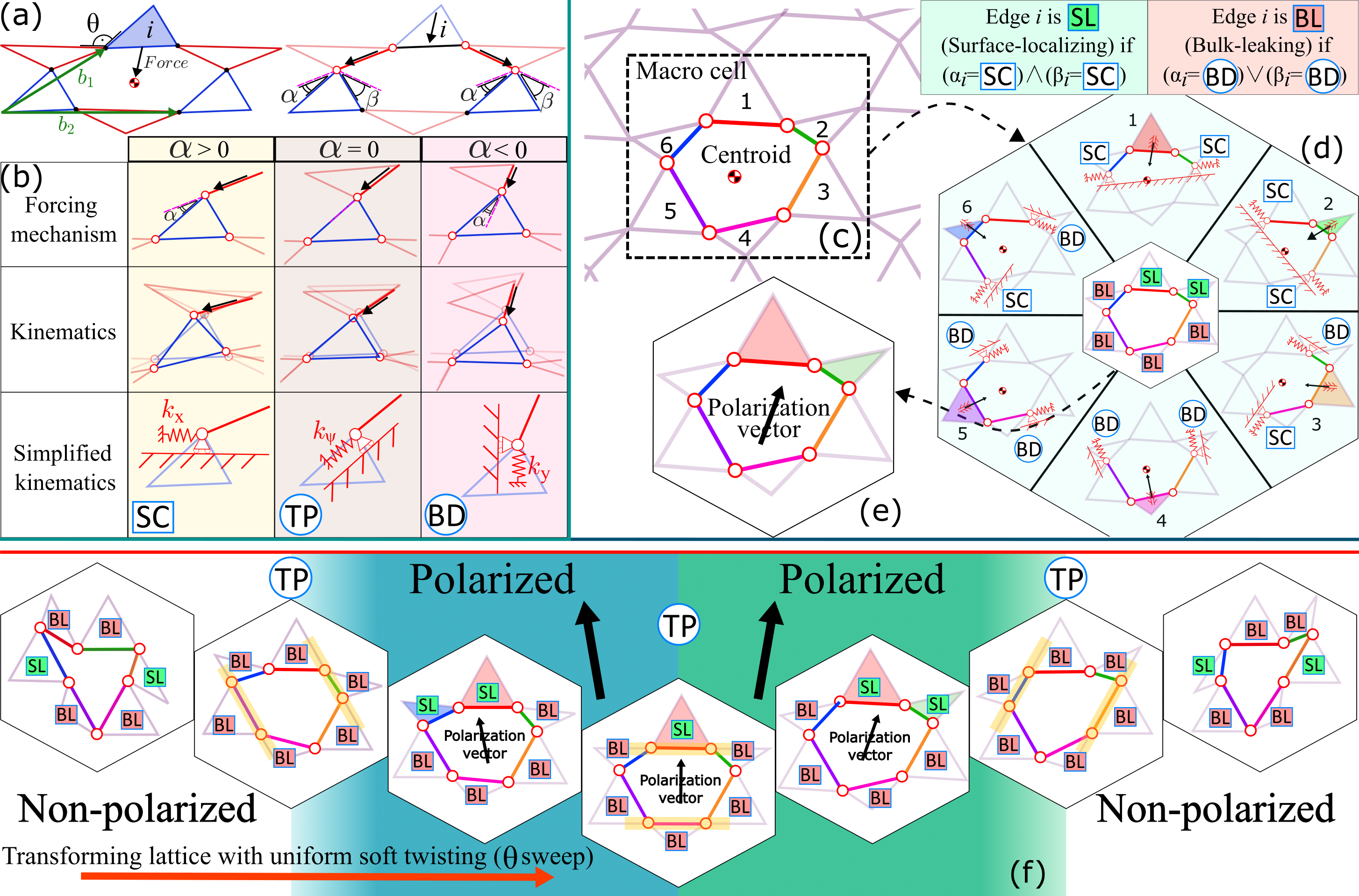}
\caption{(a) Macrocell of polarized kagome lattice encompassing six triangles and encapsulated hexagonal void, emphasizing the kinematic assumption on force direction required to filter out trivial responses. The angles $\alpha$ and $\beta$ control the kinematics of the void. (b) Taxonomy of kinematic scenarios obtained for $\alpha > 0, \, \alpha = 0, \, \alpha < 0$, and the simplified kinematic models for the corresponding hinge, labeled as surface-compliant (SC), transition point (TP) and bulk-displacing (BD), respectively. (c-e) Analysis of polarization for benchmark kagome configuration. (c) Partition of the void perimeter into six sides, identifying six trusses representative of six possible edges. (d) Workflow for characterization of the the trusses/edges: 1) the hinges of each truss are labeled as SC or BD; 2) an edge is deemed surface-localizing (SL) if both hinges are SC, and bulk-leaking (BL) if at least one hinge is BD. The central inset shows the final labeling of all the trusses/edges, this case featuring two adjacent SL edges. (e) The polarization vector points along a direction intermediate between the SL edges. (f) The method is tested against the family of kagome lattices studied in~\cite{rocklin2017transformable}, and repeated for different twist angles $\theta$ (reported in (a)) spanning the polarized and non-polarized regimes. Though the lens of the proposed method polarized configurations display one or two adjacent SL edges; non-polarized configurations display two SL edges located on opposite sides of the void; in phase transition configurations all six edges are deemed BL. The method correctly predicts the boundaries between polarized and non-polarized regions and, in the polarized range, the qualitative inclination (left- or right-leaning) of the polarization vector.}
\label{fig:intro}
\end{figure*}

The main objective of this study is to document experimentally the onset of dichotomous behavior of a structural topological kagome lattice under sharp edge loads, with and without embedded magnets. Specifically, we intend to understand whether the embedding of magnets can selectively trigger bistable configurations at the edges and determine which edge (soft or stiff) is more amenable to this transition. One of the challenges arising in designing a framework to capture this behavior experimentally is determining how to load the edges in a way that truly elicits the topological character of the bulk, instead of merely activating any trivial compliance. This issue, while fairly inconsequential for the small loads involved in linear elastic problems, becomes critical when dealing with large deformation of soft specimens \cite{widstrand2024robustness,pishvar2020soft,jolly2023soft}.
These considerations compel us to seek a deeper \textit{mechanistic} understanding of how the neighborhood of an edge cell subjected to loading deforms, and how this deformation penetrates in the bulk. Through this exercise, we
seek to build an intuitive framework to predict the polarization (or lack thereof) of a given kagome configuration only relying on geometric and kinematic considerations. 
While the proposed approach is deliberately pragmatic and does not claim to capture the topological attributes of a lattice, it does provide a simplified, yet reliable way to predict the availability of polarization. Therefore, 
it can provide a powerful complement to the formal theory of topological polarization, serving a practical purpose reminiscent of the model-free method to predict polarization recently put forward by Guzman et al.~\cite{hammer_Guzman}.

Let us consider a macrocell of kagome cells nearing an edge, consisting of one hexagonal void encapsulated by six triangles, as shown in Fig.~\ref{fig:intro}(a,c). We will show that the lattice structural properties are effectively determined by the geometric features of the void. We make the assumption that the lattice can be described as a truss, i.e., a collection of rods linked by ideal hinges. A second important phenomenological assumption is that the load is applied as a force vector passing through the mid point of the interior side of the loaded triangle (along the perimeter of the void) and directed towards the centroid 
of the macrocell, a choice of orientation that guarantees balance of angular momentum, avoiding rotations of the cell. 
This assumption is dictated by an explicit attempt to minimize possible activation of trivial edge effects, which typically involve isolated rotations of the edge triangles. 
We then proceed to select a loading triangle (labeled $i$ in Fig.~\ref{fig:intro}(a)), which automatically selects the specific edge (among the many that can be realized by severing a finite domain from an infinite lattice) along which we intend the load to be applied. 

The two vertices bounding $i$ are modeled as internal hinges, while the next pair down is taken to behave differently depending on the value of the angle $\alpha$ (on the left), or $\beta$ (on the right), between the loading rod in the upper triangle marked in red and the loaded rod in the lower triangle marked in blue. 
From this point forward, we will specialize the formulation to $\alpha$, but a similar rationale applies to 
$\beta$. 
As discussed in \cite{zhou2018topological}, the sign of 
$\alpha$ dictates the capacity of a lattice cell to transmit the load outward or downward, resulting in different effective kinematics for the truss. 
Fig.~\ref{fig:intro}(b) shows the different $\alpha$ regimes. For $\alpha > 0$, the force applied to the hinge activates mostly a lateral displacement, which we model as a horizontal roller constrained by a later spring. These simplified kinematics capture a deformation that remains mainly localized at the edge, affecting the lateral cells along the edge with negligible penetration into the bulk. We refer to this hinge condition as a \textit{surface-compliant} (SC) hinge. 
Conversely, for $\alpha < 0$, the load is transmitted mainly downward. The resulting kinematic behavior can be approximated by a constrained vertical roller. Since the deformation penetrates into the bulk, this hinge condition is defined as a \textit{bulk-displacing} (BD) hinge.
The case of $\alpha= 0$ marks an abrupt discontinuity between the two regimes. As $\alpha\to 0^{\pm}$, the sudden alignment of bonds generates a condition where forces and deformation can be fully transmitted axially along the inclined rod. This condition can be modeled as an inclined roller, allowing displacement only along the axial direction, under the constraint of a spring $k_{\psi}$ that captures the (possibly large) reaction force of a rod loaded axially. We refer to the hinge condition in this transition scenario, which signals a change in polarization and already displays the bulk-leaking attributes germane to $\alpha < 0$, 
as a \textit{transition point} (TP) hinge. 
According to this taxonomy, it is clear that only the SC hinge condition allows deformation modes that localize on the surface, i.e., floppy edge mode. While the strength of localization depends on the specific value of $\alpha$, the regime of polarization is solely controlled by its sign. 


Having defined two possible conditions for the hinges (SC for $\alpha > 0 $ and DB $\alpha \leq 0 $), the behavior of a given truss 
(i.e., of a given edge) depends on the cooperative behavior of the hinges that bound it. 
For each truss centered around triangle $i$, the deformation is \textit{surface-localized} (SL) or \textit{bulk-leaking} (BL) depending on the following condition:
\begin{equation}
i \, \: \text{is SL} \quad \iff \quad (\alpha_i = \text{SC}) \land (\beta_i = \text{SC})
\label{eq:00}
\end{equation}
\vspace{-0.3in}
\begin{equation}
i \, \: \text{is BL} \quad \iff \quad (\alpha_i = \text{BD}) \lor (\beta_i = \text{BD})
\label{eq:01}
\end{equation}
In words, the deformation is surface-localized (SL) if both bounding hinges are surface-compliant (SC). Conversely, the deformation is bulk-leaking (BL) if at least one of the two hinges is of the bulk-displacing (BD) type. 
In Fig.~\ref{fig:intro}(c-e), we apply this classification method to a test configuration, using the benchmark kagome lattice discussed in~\cite{rocklin2017transformable}, for which the polarization vector direction is known. 
Moving along the macrocell void, we recognize the truss connections corresponding to the six possible edges (Fig.~\ref{fig:intro}(d)). For each, we qualify the bounding hinges as either SC or BD, resulting in the scenarios shown in the six quadrants of Fig.~\ref{fig:intro}(d). Finally, using the criteria in eq.s~\ref{eq:00} and~\ref{eq:01}, we label the six edges (the six sides of the macrocell void) as SL or BL, as shown in the central inset of Fig.~\ref{fig:intro}(d).

\begin{figure*}[t]
\centering
\includegraphics[width = 1\textwidth]{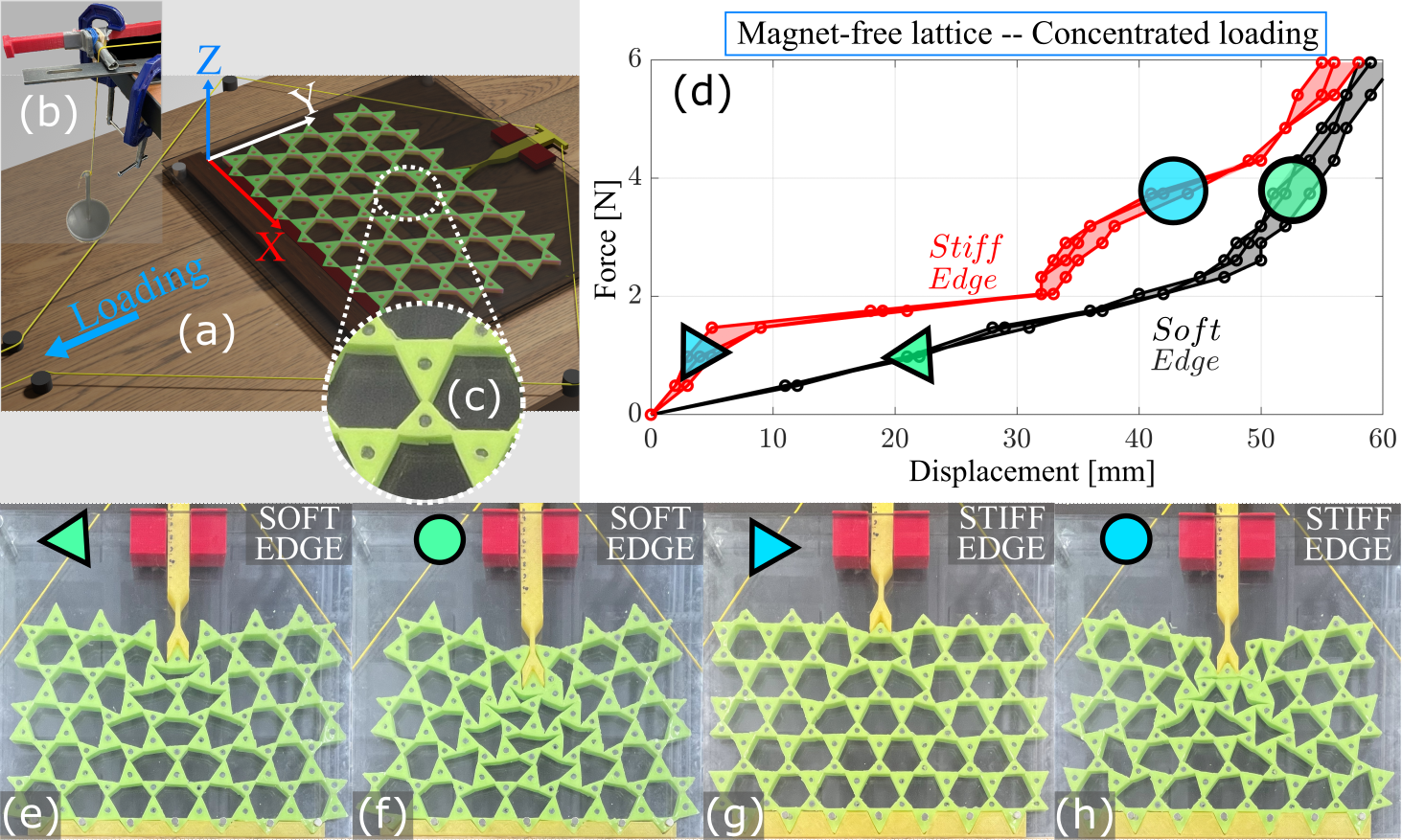}
\caption{(a,b,c) Experimental setup for force-control experiments. (a) Schematic of setup, showing specimen laid flat in a plexiglass case and pulley system to apply compressive load via dead weight, using a loading tip with V-shaped notch to prevent trivial local rotations of the loaded edge triangle. (b) Dead weight apparatus 
designed for incremental weight addition. (c) Detail of a cell of the silicone rubber prototype, highlighting finite thickness ligaments at the hinges and through holes (to be filled with plastic or magnetic fillers). (d) Experimental force-displacement curves for the lattice without magnets (purely mechanical response) under concentrated load, highlighting dichotomy between the soft and stiff edge. The markers correspond to snapshots in the (e-h) insets. (e,f) Snapshots of deformed states for load applied at the soft edge, in the linear and non-linear elastic regimes, respectively, revealing large localized deformation in the neighborhood of the tip that progressively increases with the load. (g,h) Snapshots for load applied at the stiff edge, showing stiffer response in the linear regime and sharp transition to 
localized deformation when local buckling mechanisms are triggered.}
\label{fig:ex:01}
\end{figure*}

With the ability to assign a binary label that describes the potential for surface localization to each of the possible six edges, the next challenge is to generate a single descriptor of polarization with a significance akin to that of the \textit{polarization vector} in topological theory. 
Intuition suggests that the direction of such polarization vector should point towards the edge that displays SL properties. However, there can exist situations, as the one shown in Fig.~\ref{fig:intro}(e), where more than one edge are deemed SL; note that, when this occurs, the two SL edges are adjacent. Comparing this landscape of edge properties against the outcome of formal topological analysis, which, for the configuration in Fig.~\ref{fig:intro}(e), predicts a polarization vector pointing up and leaning to the right, yields the complete assignment rule: when two SL edges are adjacent, the polarization vector is directed along an intermediate direction, with the exact angle left indeterminate, but surely dependent on the unknown ratio that relates the strength of localization of each edge.\\
The robustness of the method is showcased in Fig.~\ref{fig:intro}(f), in which we sweep the configuration space of our selected kagome family by varying the twist angle, mimicking the sweep performed in~\cite{rocklin2017transformable} to construct the polarization phase diagram. We can see, that, for a wide range of twist angles in the central portion of the sweep axis, the computed landscapes and the SL criterion predict correctly the direction and the inclination of the polarization vector, including the transition from left-leaning (blue-shaded region) to right-leaning (green-shaded region) occurring at the configuration featuring perfectly horizontal bonds. Interestingly, the TP conditions obtained when bonds align to form straight fibers in the lattice are accurate predictors of phase transitions from polarized to non-polarized. In our framework, their transition character is marked by the fact that all the edges are deemed BL, and none exhibits SL character. This implies that, at this point in the sweep, all the edges stop enjoying the excess of floppiness associated with the floppy edges of polarized configurations. 
Finally, we can consider configurations located at the two extremes of the twist angle sweep. Here, the macrocells feature again two SL edges, but these are now located on opposite sides of the void, offering conflicting characterization of the floppiness landscape. This contradiction results in an indeterminacy to deem a side floppier than the other. These are in fact configurations that are identified as non-polarized (i.e. with a null polarization vector) by formal topological analysis. This symmetric situation can be interpreted as resulting from the migration of a floppy mode from the floppy edge to the rigid one, which reestablishes the edge balance of conventional, non-polarized media. We can finally test the method against a special class of twisted kagome lattices, which feature two identical equilateral triangles rotated by a twist angle angle $\theta$ and are non-polarized for any $\theta$. It is easy to verify that the method properly captures the lack of polarization, which manifests in this case as a landscape of triangles featuring only BL edges.

\section{Experiments on soft lattices with magnets-enabled bistability} \label{sec:experiments}

With a rationale in hand for the selection of the most appropriate loading protocol, we are ready to test the nonlinear regime of deformation of \textit{soft} polarized kagome lattices. We first characterize the behavior of a \textit{purely mechanical} specimen, to assess the degree of response asymmetry observable between soft and stiff edges under finite deformation. We then shift our focus on the response of a \textit{magneto-mechanical} lattice enabled by the introduction of magnets.

Our specimen is a structural kagome lattice with unit cell shown in Fig.~\ref{fig:ex:01}(c) featuring a pair or large and small isosceles triangles. This configuration has been shown to feature a marked stiffness dichotomy between opposite edges \cite{zunker2021soft} that is robust even in realistic structural lattice conditions. 
The lattice is cast from silicone rubber (Zhermack Elite Double 32) following a popular fabrication protocol for soft metamaterials prototyping, summarized in the Materials and Methods section. The triangles feature through holes at their centroids. The holes, designed to eventually host magnets (as discussed later in the manuscript) are temporarily filled with plastic (PLA) cylinders with the same radii of the magnets to provide bulk stiffness to the triangles. 
We custom design a dead-weight loading device, schematically depicted in Fig.~\ref{fig:ex:01} (a), which allows applying precise static loads through a simple table-top apparatus. The deliberately low-tech features of the apparatus make it highly portable, inexpensive to assemble and easy to operate, and ultimately ideal for broad use across tasks and lab environments, without the need for prohibitive sophisticated equipment. The lattice is inserted in the sleeve of a Plexiglas case, which is laid flat on a table. 
To minimize friction between the case and the lattice, which would interfere with the lattice deformation, the cylinders are allowed to protrude slightly out-of-plane so that they slide against the plexiglass, preventing direct interaction between the silicone and the case.  The load is applied through a 3D-printed loading rod, with tip designed with a V-notch to engage an edge triangle in a way that prevents the triangle from rotating as the load is applied in order to minimize trivial edge deformation associated with local rotations. 
The rod is constrained to slide along a guiding sleeve to maintain the direction of loading at all stages, according to the kinematic assumptions of Fig.~\ref{fig:intro}, 
and the deflection of the edge 
can be agilely tracked through markings on the loading rod. %


\begin{figure}
\centering
\includegraphics[width = 0.475\textwidth]{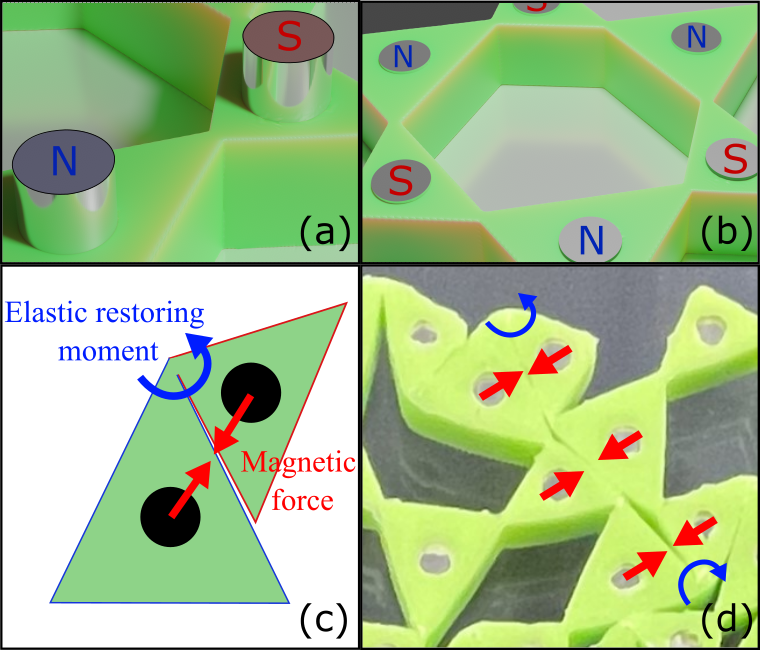}
\caption{(a-b) Renders of the magnets insertion. (b) Magnets positioning in the lattice according to a North-South pole alternation between neighboring sites. 
(c) Schematic showing competing effects towards moments balance of magnetic attractive forces and elastic reactions forces. (d) Example of bistable configuration triggered by magnets in a small physical prototype with the same geometric, material and magnetic characteristics of the tested lattices.}
\label{fig:ex:magnets}
\end{figure}

\begin{figure*}[t]
\centering
\includegraphics[width = 1\textwidth]{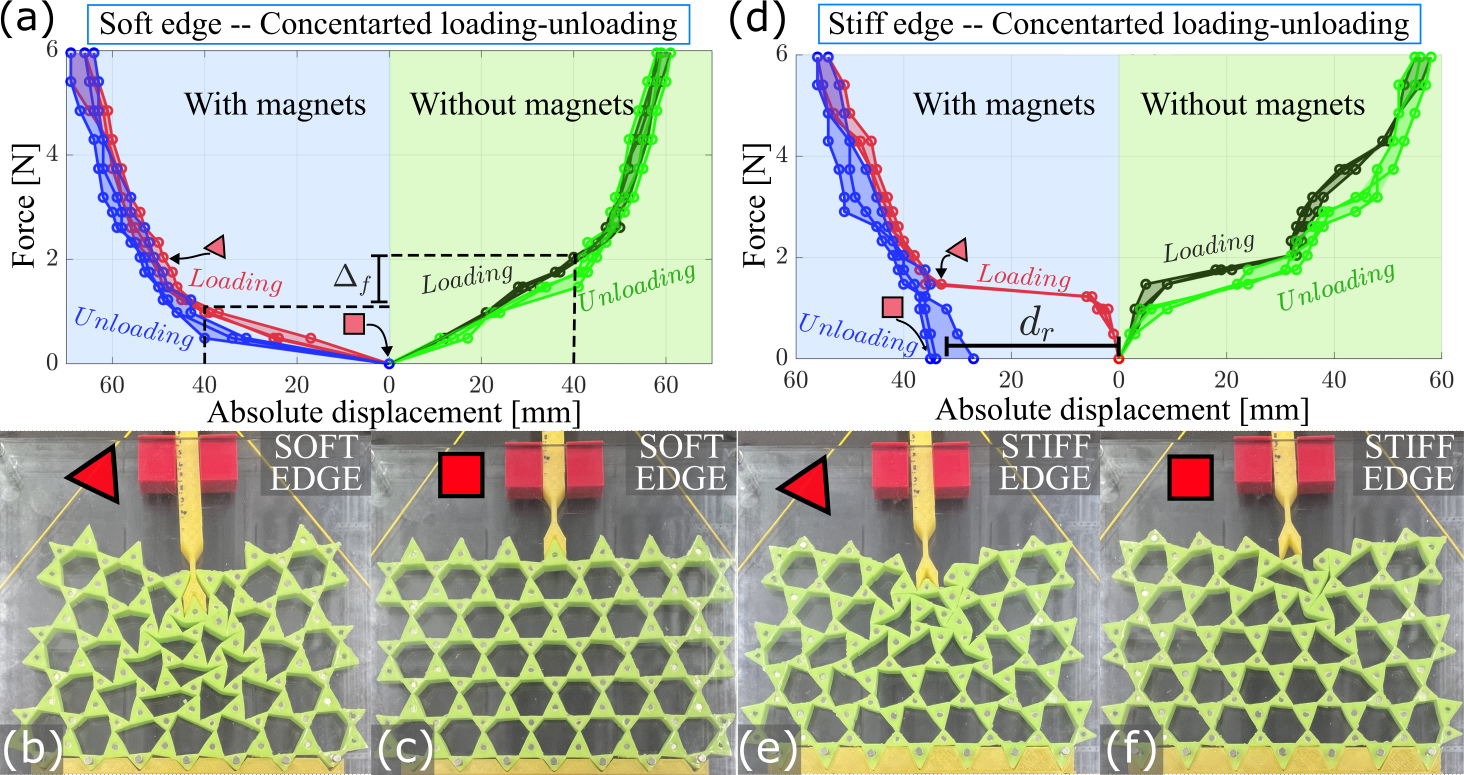}
\caption{Loading and unloading tests of polarized soft lattices with magnets, endowed with bistable mechanisms. (a-c) Results for loading at the soft edge. (a) Loading and unloading force-displacement curves (shaded blue background), compared against the reference case without magnets (shaded green background) and showing additional softening introduced by the magnets showcased in the form of $\Delta_f$: force delta needed to achieve similar displacement of $\approx 40 \textrm{mm}$. The superimposed curves for loading and unloading are almost identical, suggesting elastic behavior throughout the process. (b) Snapshot of deformed lattice, with cell pattern resembling the case without magnets. (c) snapshot of lattice upon unloading, showing the undeformed configuration fully recovered and confirming elastic behavior for loading at the soft edge. (d-f) Results for loading at the stiff edge. (d) Loading and unloading force-displacement curves with and without magnets; the loading curves are qualitatively matching, notwithstanding a quantitative shift due to softening; the unloading curve for the case with magnets reveals activation of irreversible deformation marked with residual edge deflection $d_r$ upon unloading. (e) Snapshot of deformed state. (f) Snapshot of lattice at the end of the unloading stage, showing residual deformation and edge deflection.}
\label{fig:ex:03}
\end{figure*}

Fig.~\ref{fig:ex:01}(d) reports the force-displacement curves measured during the loading process for loads applied to the soft (black curve) and stiff (red curve) edges, with snapshots of the deformed specimens in the Fig.~\ref{fig:ex:01}(e-h) insets for key stages of the process denoted by the markers. The most notable thing is the dichotomy in stiffness between the edges, inferred from the slope of the curves, with the edge predicted to be floppy 
displaying indeed significantly higher compliance. The dichotomy is evident comparing the insets (e) and (g), marked by triangular markers, showing the stiff and soft edge response under the same force ($\approx 1 \, \textrm{N}$). This result is in agreement with polarization theory, of which it represents an important extension into the finite deformation regime, and consistent with the observations in~\cite{pishvar2020soft}. The second -- and less obvious -- observation regards the large gap between the ranges of displacements over which the response can be considered linear elastic, ranging from $\approx 5 \, \textrm{mm}$ for the stiff edge to $\approx 45 \, \textrm{mm}$ for the soft one, corresponding to edge deflections of $\approx 2.75\, \%$ and $\approx 24.73 \, \%$ of the specimen height, respectively. Beyond a certain value of force ($2 \, \textrm{N}$ and $\approx 2.5 \, \textrm{N}$ for the stiff and soft edge, respectively) and beyond a certain value of displacement ($\approx 35 \, \textrm{mm}$ and $\approx 50 \, \textrm{mm}$), 
the curves steepen significantly. This stiffening occurs when the deformation is so pronounced that the triangles come into contact, approaching the behavior of a continuously solid silicone slab.

\begin{figure}
\centering
\includegraphics[width = 0.5\textwidth]{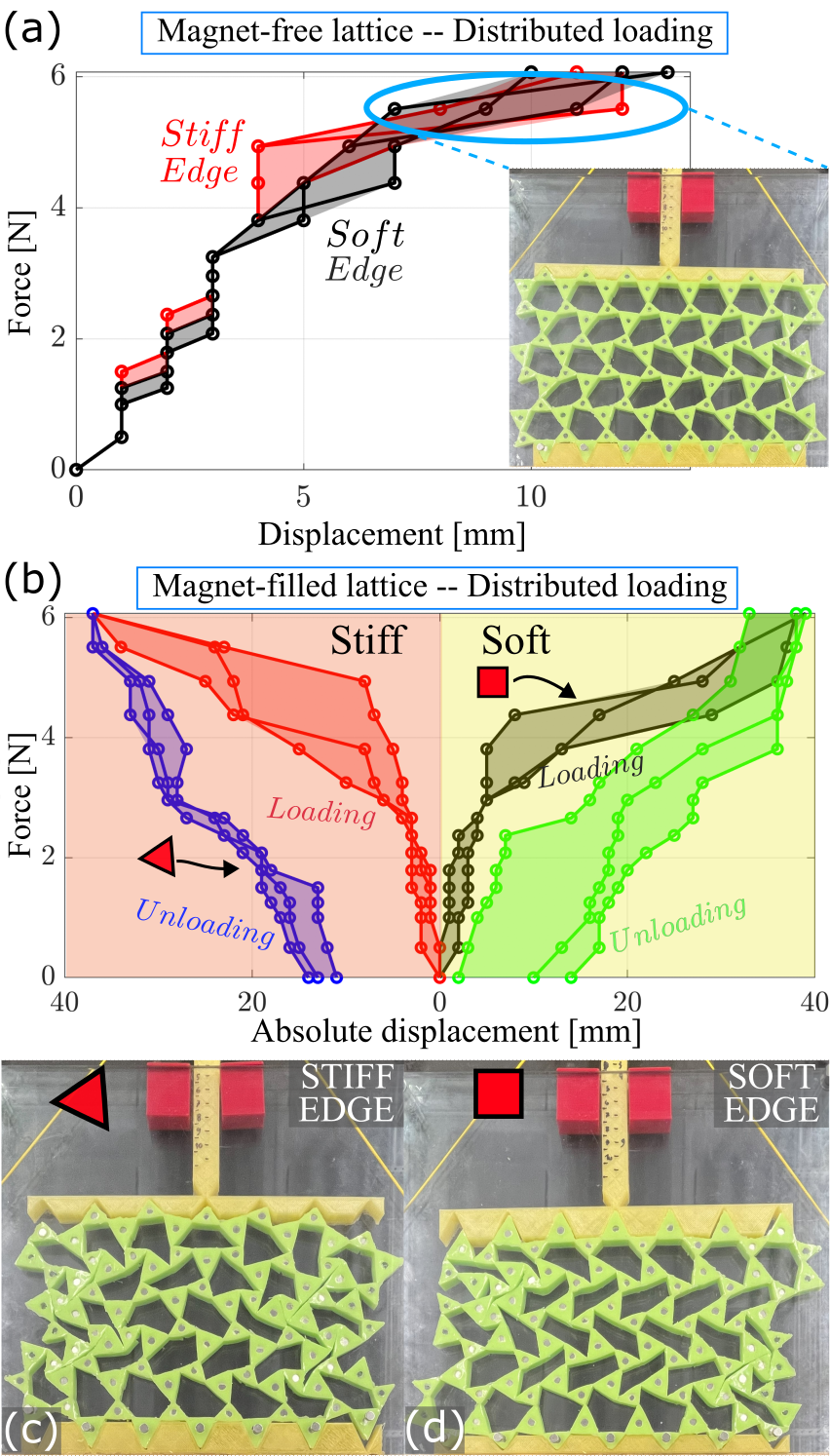}
\caption{(a) Force-displacement curves of purely mechanical lattice under distributed loading, showing no appreciable differences between soft and stiff edge, in contrast with the concentrated load case of Fig.~\ref{fig:ex:01}. (b) Loading and unloading force-displacement curves for lattice with magnets under distributed loading condition, comparing stiff edge loading (shaded red background) vs. soft edge loading (shaded yellow background). (c,d) Snapshots of the lattice under large deformation applied at the stiff and soft edge, respectively, with both cases showing extensive zones experiencing bistability, albeit with different random patterns of buckled cells.}
\label{fig:ex:05}
\end{figure}


We now proceed to characterize the magneto-mechanical response of a lattice endowed with magnets. The magnets are inserted in the holes as shown in Fig.~\ref{fig:ex:magnets}(a), and arranged in the lattice such that two magnets in adjacent triangles feature opposite North-South poling, as shown in Fig.~\ref{fig:ex:magnets}(b). The triangles along the free left/right edges are left without magnets; this is because, at the edges, where the lattice coordination is lower, the restoring elastic forces are low and any magnetic forces would result in local (trivial) instabilities even under small loads. 
For an undeformed cell, the direction of the attractive force between two neighboring magnets 
passes through the hinge connecting the triangles, thus exerting a negligible moment about the hinge. In contrast, when the cell is deformed, the magnetic force exerts a couple that tends to drive the magnets closer to each other (see Fig.~\ref{fig:ex:magnets}(c)), counteracted by the restoring moment due to the elastic strains stored in the bent hinge.  The interplay between these opposing moments dictates the stable equilibrium configurations onto which the lattice settles. Specifically, the lattice displays bistable behavior: under small loads, the lattice is in a 
configuration where 
the magnets are maximally distant and the cell voids retain their hexagonal shapes; when the load exceeds a critical value --- and certain triangles rotate beyond a critical twist --- the moments due to the magnets overcome the restoring ones due to elasticity and the lattice snaps to a stable configuration in which the edges of the triangles touch and the magnets reach their minimum allowable distance. This bistable behavior is illustrated on a sample of cells with magnets in Fig.~\ref{fig:ex:magnets}(d).

We repeat the tests performed on the purely mechanical lattice (i.e. concentrated and distributed loading of soft and stiff edges) on the magneto-mechanical lattice. Importantly, here we explicitly track the history of the \textit{unloading path} with the objective of detecting any 
residual deformation developed during the activation of a bistable configuration. Note that we also recorded the unloading path for the mechanical lattice, but did not detect any appreciable differences with respect to the loading path, indicating that, in that case, the behavior was perfectly elastic.
The results for loads applied at the soft and stiff edge are reported in Fig.~\ref{fig:ex:03}(a,b,c) and (d,e,f), respectively. For each case, we report the force-displacement curves (Fig.~\ref{fig:ex:03}(a) and (d)) depicted next to the reference curves without magnets using specular 
plots to appreciate symmetries, or lack thereof, between the scenarios. 

The curves for the 
soft edge (Fig.~\ref{fig:ex:03}(a,b,c) leads to two observations. First, while the loading trend is almost identical with and without magnets, the presence of the magnets does manifest as an overall softening of the lattice, denoted by a shallower curve, which can be explained by noting that the pairs of attractive magnets, when brought closer via deformation, add a discrete force network that promotes deformation. This is marked by a reduction $\Delta_f \approx 50 \%$ of the force required to establish a deflection of $\approx 40 \textrm{mm}$. The inset of Fig.~\ref{fig:ex:03}(b) for the stage denoted by the triangular marker reveals a deformed configuration that is, for all intents and purposes, analogous to that without magnets in Fig.~\ref{fig:ex:01}(f). The second observation is that, by enlarge, the unloading curves overlap with the loading ones, indicating that the behavior remains perfectly elastic 
and the original configuration is fully recovered upon loading, as captured in the inset of Fig.~\ref{fig:ex:03}(c) for the stage marked by the square marker. The response for 
the stiff edge (Fig.~\ref{fig:ex:03}(d,e,f)) is markedly different. Again, the loading phase is qualitatively similar to the case without magnets, but features a much softer response (for reasons analogous to those invoked for the soft edge).
The unloading path, instead, deviates significantly, quantitatively and qualitatively, from the magnet-free case. Upon full unloading, the undeformed configuration is never recovered, as shown in Fig.~\ref{fig:ex:03}(d,e,f) and displays 
a residual 
deformation quantified by the residual displacement $d_r$ of the edge loading point, shown Fig.~\ref{fig:ex:03}(d), of $\approx 30 \textrm{mm}$, corresponding to $\approx 50 \%$ of the total indentation depth.

In summary, the experiments reveal that the bistable attributes of the lattice are tapped into only when loading the stiff edge (Fig.~\ref{fig:ex:03}(d,e,f). This seems surprising, given that the deformation recorded at the soft edge is more pronounced. However, visual inspection of the cells patterns under loading indicate that, at the stiff edge, large loads induce local buckling episodes that cause large rotations and eventually engage more magnet pairs, bringing them into closer proximity. On the other hand, on the soft edge, a vertically applied load does not establish the rotations needed to tap into the bistable configuration. In essence, the asymmetry of stiffness between the edges results in a different promotion of the conditions for the activation of bistability, which in turns further accentuates the asymmetry of the mechanical response, thus extremizing the dichotomy between the edges. The dichotomy is \textit{quantitative}, as it involves an exacerbation of the difference in linear stiffness between the edges observed for the purely mechanical lattice. 
It is also \textit{qualitative}, because the activation of irreversible mechanisms is available \textit{selectively} on the stiff edge only. Moreover, the dichotomy is counter-intuitively opposite to that observed for the purely mechanical lattice, in the sense that, here, it is the stiff edge that, by experiencing irreversible reconfiguration, carries the most drastic signature of deformation. Finally, an important difference emerges between the regular kagome 
studied in~\cite{Ruzzene_mag} and the configuration considered in this work. 
In the regular kagome case, the activation of bistability induces a cascade of rotations that propagates as a wave and rapidly affects the whole domain, reconfiguring the \textit{entire} lattice into its most compact state, with no intercell voids between the triangles. Here, in contrast, the geometric incompatibility between neighboring triangles, due to the length mismatch of their sides, makes the reconfiguration self-limiting. As a triangle engages its neighbor in a way that does not completely fill the intercell void, it is locked in position, generating a local defect in the periodic pattern that precludes further changes in configuration.\\
Through a simple change in the setup, where we replace the loading tip with a solid slab fitting the uneven profile of the entire edge, we can replicate the tests under a distributed load for the case of purely mechanical lattice (Fig.~\ref{fig:ex:05}(a)). Clearly, the two curves for the stiff and soft edges do not show any appreciable difference, confirming that the edge dichotomy emerges under concentrated loads and is bypassed by distributed ones. 
We then complete our experimental set by repeating the tests with magnets under distributed loading. For the loading path, we again do not observe appreciable differences between the soft and stiff edges. In both cases, the magnets introduce a general softening of the lattice, inferable by comparing the results of Fig.~\ref{fig:ex:05}(a) and Fig.~\ref{fig:ex:05}(b): with magnets, a load of $6 \, \textrm{N}$ brings about a displacement of $\approx 30 \, \textrm{mm}$, compared to $\approx 10 \, \textrm{mm}$ for the purely mechanical case. For the unloading path, we do observe some residual displacement upon unloading from either side, provided that the loading phase has reached a sufficiently advanced level of deformation, with the effect being more pronounced for the stiff edge. In this case, both edges lead to formation of clusters of buckled cells. The number and location of these clusters appears to be obey random processes, likely driven by imperfections in the lattice and slight asymmetries in the load. This results in the establishment of arbitrary spatial patterns of bistable cell patches which, at the moment, elude a precise rationale that would satisfactorily explain the difference between stiff and soft edge. It is worth reporting that, in all experiments for which residual deformation is achieved, the original underformed configuration can be recovered by inducing a sudden motion of the entire structure, e.g. via a relatively low-amplitude impulsive load (impact). This relaxation is due to the relative low energy barrier of the bistable well granted by the magnetic interactions. These energy barriers can be easily overcome by the kinetic energy imparted to the system.

\section{Conclusions} \label{sec:Conclusions}

We have illustrated the emerging mechanical response of soft magneto-mechanical Maxwell metamaterials from the cooperative interplay between mechanical edge asymmetry (rooted in topological polarization), finite deformation (due to softness of rubber as a fabrication material), and bistability (enabled by the magnets). We have shown that the polarization drives the establishment of bistable mechanisms (triggered under large deformation) in different ways on the soft and stiff edge, which in turn produces different cell reconfiguration patterns. We have documented a complex augmentation of the inherent asymmetry of purely mechanical lattice, whereby the presence of the magnets can either extremize or counteract the asymmetry due to polarization, according to the regime of deformation.

The ability to store energy asymmetrically and the availability of a mechanism for its quick release endow magneto-mechanical polarized lattices with interesting impact protection and energy harvesting capabilities. Three ingredients contribute collaboratively to the establishment of these effects. Specifically, the ability to retain polarized behavior in the large deformation regime translates into an ability to accommodate extreme local deformation (e.g. under indenting loads) and effectively store potential energy, without requiring excessive bulk softness of the entire lattice, thus working with a structural metamaterial that features satisfactory load-bearing capabilities. 
The second ingredient is the bistability granted by the magnetic force network, which enables the ability to retain deformation and energy even upon unloading, playing the role of a mechanical storage system for the harvesting system. Finally, the availability of a method for on-demand quick release of stored energy through an excitation that is modaly distinct from the load path through which energy is stored adds a layer of flexibility 
to the energy harvesting strategy.

\section{Materials and Methods}

The specimen is created using the following method. Initially, a negative mold is fabricated using 3D printing with PLA, featuring pre-designed pillars to accommodate the magnets' holes. Subsequently, the silicone rubber Zhermack Elite Double 32 mixture is poured into the mold, after undergoing a degassing process. When the material sets, the lattice structure is carefully extracted, and the magnets are precisely inserted.
This production procedure mirrors the approach detailed in \cite{widstrand2024robustness}, enabling inference of material behavior and parameters from that study. Hinge thickness was empirically validated to meet the requisite compliant characteristics for establishing the bistable potential well.
A bespoke pulley testing setup, featuring a maximum load cell capacity of 8 N, was employed to assess the load-displacement characteristics of the structural lattices.
The lattices were constrained during loading procedures utilizing both 3D printed holders and magnets positioned on the last row, achieving a pinned boundary condition on the bottom of the lattice.
The magnets integrated into the lattice are NdFeB grade N52, measuring 9.5 mm in height and 6.3 mm in diameter.


\section*{Acknowledgements} \label{sec:acknowledgements}

L. Iorio and R. Ardito acknowledge the support of the H2020 FET-proactive Metamaterial Enabled Vibration Energy Harvesting (MetaVEH) project under Grant Agreement No. 952039. S. Gonella acknowledges the support of the National Science Foundation (grant CMMI - 2027000).




\bibliography{bibliography}


\end{document}


\title{Supplementary material}

\section{Fabrication}
The specimen is fabricated through the following procedure. First a mold is 3D printed in PLA. The mold is the negative of the final structure and already features pillars that generate holes that will host the magnets.
Then the silicone rubber Zhermack Elite Double 32 mixture is cast into the mold. The lattice is then extracted after setting and then the magnets are carefully introduced.
The magnets introduced in the lattice are NdFeB grade N52, of height 9.5 mm and diameter 6.3 mm.
The production procedure is the same one adopted in \cite{widstrand2024robustness}, consequentially the material behaviour and parameters can be inferred from that work.


\begin{figure}
\centering
\includegraphics[width = 0.8\textwidth]{sections/Figura_experimental_setuo.png}
\caption{Experimental setup. (a) Pulley system that hosts increasing weights. (b) Work table holding the Plexiglas case that hosts the rubber lattice; lateral pulley systems ensure the load transmission from the weight to the indenter.}
\label{fig:exp}
\end{figure}

A  custom-  made  tensile  testing  setup with a 1 N load cell and high- precision moving stages were used to char-acterize the load- displacement responses of the magnetoelastomeric beams. The beams were fixed in custom 3D- printed frames with different angles which were then placed on the xy- stage.  The z-  stage was moved at a speed of 0.1 mm s−1 and the load- displacement data were recorded at a frequency of 1,000 Hz.